\documentclass[preprint,12pt]{elsarticle}

\usepackage{amssymb}

\journal{Electrochemistry Communications}

\begin{document}

\begin{frontmatter}



\title{Poisson-Nernst-Planck Model of Bipolar Nanofluidic Diode Based on Bulletlike Nanopore}


\author[bnu]{Li-Jian Qu}
\author[bju]{Xinghua Zhang}
\author[bnu]{Jie Fu}
\author[bnuchem]{Lin Li}
\author[bnu]{Dadong Yan\corref{cor1}}
\ead{yandd@bnu.edu.cn}
\cortext[cor1]{Corresponding author. Tel.: +86 10 58807604; fax: +86 10 58807604.}

\address[bnu]{Department of Physics, Beijing Normal University}
\address[bju]{Department of Physics, Beijing Jiaotong University}
\address[bnuchem]{College of Chemistry, Beijing Normal University}

\begin{abstract}
  Bipolar nanofluidic diode is based on nanopore with positive and negative surface charges separated by a junction. This paper investigates the effects of the pore structure, taking the bullet-like pore as an example, on the ion current rectification. The Poisson-Nernst-Planck Modelings show that the ion current rectification behavior can be greatly influenced by the shape of the pore. The bipolar nanofluidic diode with more tapered tip has significantly higher ion current rectification degree. The modelling results indicate that special design of the nanopore is necessary for the performance of the bipolar nanofluidic diode.

\end{abstract}


\end{frontmatter}


\section{Introduction}
\label{}
Synthetic nanopores and nanochannels have excellent performance in manipulating mass\cite{bhatia2006transport} and charge\cite{daiguji2010ion,cheng2010nanofluidic,hou2011biomimetic,gracheva2006simulation} transport on nanoscales. The nanopores can be used to build nanofluidic devices\cite{cheng2010nanofluidic,siwy2010engineered} controlling ion transport in solution similar to semiconductor devices controlling the transport of electrons. Nanopore has been used to build various nanofluidic logic devices. The nanofluidic diode, utilized for rectification of ionic transport, can be built on nanopores with the inner surface coated with charge. The surface charges may be of the same sign\cite{siwy2006ion}. The ion current rectification will be enhanced when a half of the channel is coated with opposite sign or electrically neutral\cite{nguyen2010comparison}, which are called bipolar and unipolar ionic diode respectively. Siwy's group\cite{kalman2008nanofluidic} fabricated bipolar transistors by combining two bipolar ionic diodes and forming a PNP junction in the nanopore. Field effect transistor (FET) has been built on nanopore with an additional electrode on the wall\cite{karnik2005electrostatic}. A field-effect reconfigurable nanofluidic diode has been proposed by introducing an asymmetric field effect along the nanochannel\cite{guan2011field}. Cheng and Guo demonstrated an ionic triode composed of positive-charged alumina and negative-charged silica nanochannels\cite{cheng2009ionic}.

For the nanopores asymmetric with respect to wall charge distribution or/and pore shape, a nonlinear diodelike current-voltage response, ion current rectification (ICR), can be observed. The surface charge provides an convenient way to control ions' flow in nanopore. Nanopores with its two halves of the inner surface with opposite charges can work as a nanofluidic diode similar to the p-n semiconductor diode\cite{vlassiouk2007nanofluidic}. This bipolar nanofluidic diode produces higher ion current rectification than the nanofluidic diode based on nanopores coated with charge of only one sign\cite{nguyen2010comparison,constantin2007poisson}. Kumar and his colleagues performed systematic theoretical modellings on the fluidic bipolar nanopore. Their theoretical calculations  predict that sharp junction is not necessary for bipolar nanofluidic diode to rectify the ion current\cite{singh2011ion,singh2011effectof}. The ICR degree decreases with the diameter of the nanopore and the ion concentration\cite{singh2011effect}.

Previous studies on bipolar nanofluidic diodes paid little attention to the effects of the pore shape. Deep understanding on the relation between the nanopore properties and the pore shape helps to fabricate nanopores with optimal structural asymmetry. In addition, the control of the tip shape is also a key issue for the applications in molecular separation and sensing\cite{lee2004electrophoretic}. In the present paper, we explore the effects of tip shape on the rectification properties of the bipolar nanofluidic diode, taking the bullet-like nanopore as an example.

\section{Theoretical Model}

\begin{figure}
\centering
\includegraphics[scale=0.5]{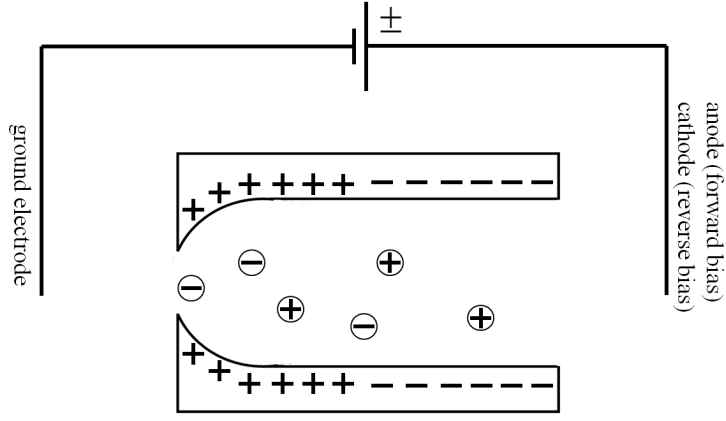}
\caption{Schematic illustration of the bipolar fluidic diode based on a bulletlike nanopore. }
\label{schematic}
\end{figure}

Figure \ref{schematic} shows schematically a bipolar fluidic diode based on a bulletlike nanopore. The nanopore radius can be described by the following equation, introduced by Ramirez, et al\cite{ramirez2008pore,cervera2011asymmetric},
\begin{equation}\label{radius}
  R(z)=\frac{R(L)-R(0)\exp(-L/h)-[R(L)-R(0)]\exp[-(z/L)(L/h)]}{1-\exp(-L/h)}
\end{equation}
where $R(0)$ and $R(L)$ are the radii at the left and right pore ends, respectively, $L$ is the length of the pore, and $L/h$ controls the shape of the nanopore.

The density of the surface charge is assumed to be described by the following function\cite{singh2011ion,singh2011effectof}
\begin{equation}\label{surfcharge}
  \sigma(z)=\sigma_0 \left (1-\frac{2}{1+\exp[-k(z-z_0)]}\right )
\end{equation}
where $\sigma_0$ is the surface charge density in the limit $z \to 0$, $z_0$ is the position where $\sigma(z)=0$, and $k$ characterizes the sharpness of the junction.

The ion transport through the nanopore is governed by the Poisson-Nernst-Planck equations
\begin{equation}\label{Poisson}
  \nabla^2\phi(\vec{r}) = -\frac{F^2}{\varepsilon RT} \left( \sum_i \nu_i c_i(\vec{r}) + \delta[r-R(z)]\sigma(z)/F \right)
\end{equation}
\begin{equation}\label{NP}
  \vec{J}_i(\vec{r}) = -D_i \left [ \nabla c_i(\vec{r}) + \nu_i c_i(\vec{r}) \nabla \phi(\vec{r}) \right ]
\end{equation}
and the continuity equation
\begin{equation}\label{continuity}
  \nabla \cdot \vec{J}_i(\vec{r}) = 0
\end{equation}
where $\vec{J}_i(\vec{r})$, $D_i$ and $\nu_i$ are the molar ion flux density, the diffusion coefficient and the valence of ion $i$; $F$, $R$ and $T$ are the Faraday constant, gas constant and the absolute temperature; $\phi$ and $\varepsilon$ are the local electric potential in $RT/F$ units and the electrical permittivity of the solution.

In the present study, we consider the situation of long and narrow nanopore, so the electro-osmotic effects are ignored. To simplify the problem, we further assume that the ion fluxes have only an axial component,
\begin{equation}
  \vec{J}_i(\vec{r}) = J_i \hat{z}
\end{equation}
where $\hat{z}$ is the unit vector along the axial direction. The ion concentrations and electric potential are assume to be constant across each cross section along the nanopore axis, $c(\vec{r})=c(z)$, $\phi(\vec{r})=\phi(z)$. Theoretical modellings on bipolar nanofluidic diode based on conical nanopore\cite{constantin2007poisson} give us confidence on these approximations.

Following the procedure developed by Kosinska et al\cite{kosinska2008rectification}, the 3D PNP model can be reduced to be 1D PNP model. Eq.(\ref{Poisson}) transforms into
\begin{equation}\label{Poissontrans}
  \frac{{\rm d}^2\phi}{{\rm d} z^2}+\frac{{\rm d} A(z)}{{\rm d} z}\frac{{\rm d} \phi}{{\rm d} z}+\frac{F^2}{\varepsilon RT}\left( \frac{2\sigma(z)}{F R(z)}+\sum_i\nu_i c_i \right)=0
\end{equation}
Combining Equations (\ref{NP}) and (\ref{continuity}), we get the equation for ion currents $I_i$:
\begin{equation}\label{NPtrans}
  \frac{I_i}{\nu_i F D_i A(z)} + \nu_ic_i\frac{{\rm d} \phi}{{\rm d} z}+\frac{{\rm d} c_i}{{\rm d} z}=0
\end{equation}
where $A(z)=\pi R^2(z)$ is the variable cross section area. The total ion current is
\begin{equation}
  I=\sum_i\nu_iI_i
\end{equation}
We adopt the Donnan equilibrium boundary conditions at the ends of the nanopore \cite{cervera2006ionic,constantin2007poisson}, given as
\begin{equation}\label{cbound}
  c_i(z_j)=-\frac{\nu_i}{FR(z_j)}+\sqrt{\left(\frac{\sigma(z_j)}{FR(z_j)^2+c_{0i,j}^2}\right)}
\end{equation}
\begin{equation}\label{potentialbound}
  \phi(z_j)=\phi_j-\frac{1}{\nu_i}\ln\frac{c_i(z_j)}{c_{0i,j}}
\end{equation}
where $c_{0i,j}$ denotes the bulk concentration of ion $i$ and $j=L, R$ represents the left and right end. $z_j$ denotes the pore borders, $z_L=0$ and $z_R=L$. The equations (\ref{Poissontrans}) and (\ref{NPtrans}) together with the boundary conditions (\ref{cbound}) and (\ref{potentialbound}) form the base for theoretical modelling.

\section{Results and Discussion}

In our calculations, the parameters are taken as follows if not specified. The length of the nanopore is $L=12000nm$. The nanopore radii at the left and right ends are $R(z_L)=3nm$ and $R(z_R)=300nm$ respectively. The ions are monovalent, i.e., $\nu_{\pm}=\pm 1$. The bulk concentration is set at $c_{0i,L}=c_{0i,R}=0.1M$. The diffusion coefficients are taken as constant value, i.e., $D_{+}=D_{-}=2.0\times 10^{9}nm^2/s$. The system is at room temperature $T=298K$. For parameters in the surface charge distribution of Eq.(\ref{surfcharge}), we set $\sigma_0=1.0$ and $k=30$. In our calculations, the electrode on the pore tip side is kept grounded, i.e., $\phi_L=0V$.
\begin{figure}
\centering
\begin{minipage}[t]{0.3\textwidth}
    \centering
    \includegraphics[scale=0.25]{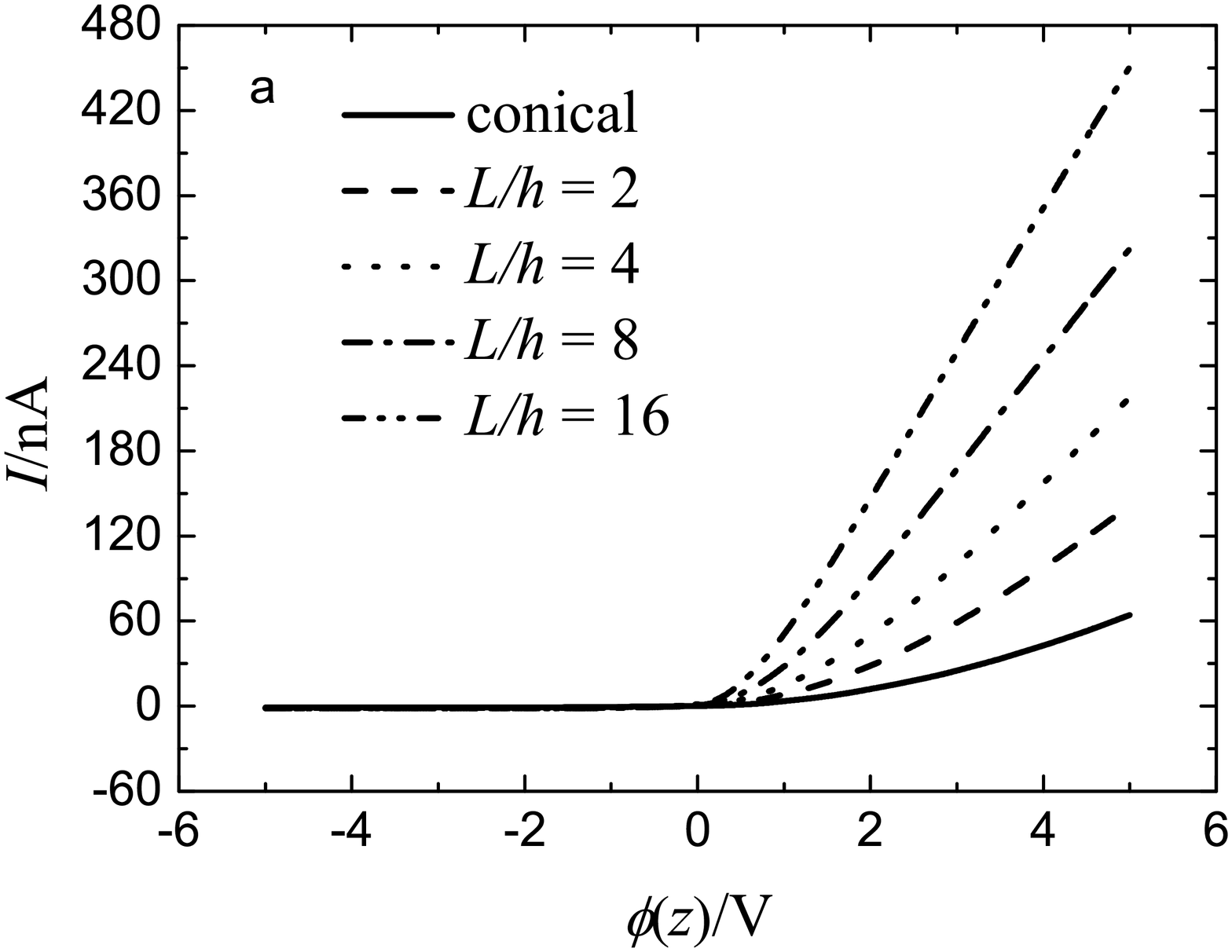}
\end{minipage}
\hspace{60pt}
\begin{minipage}[t]{0.3\textwidth}
    \centering
    \includegraphics[scale=0.25]{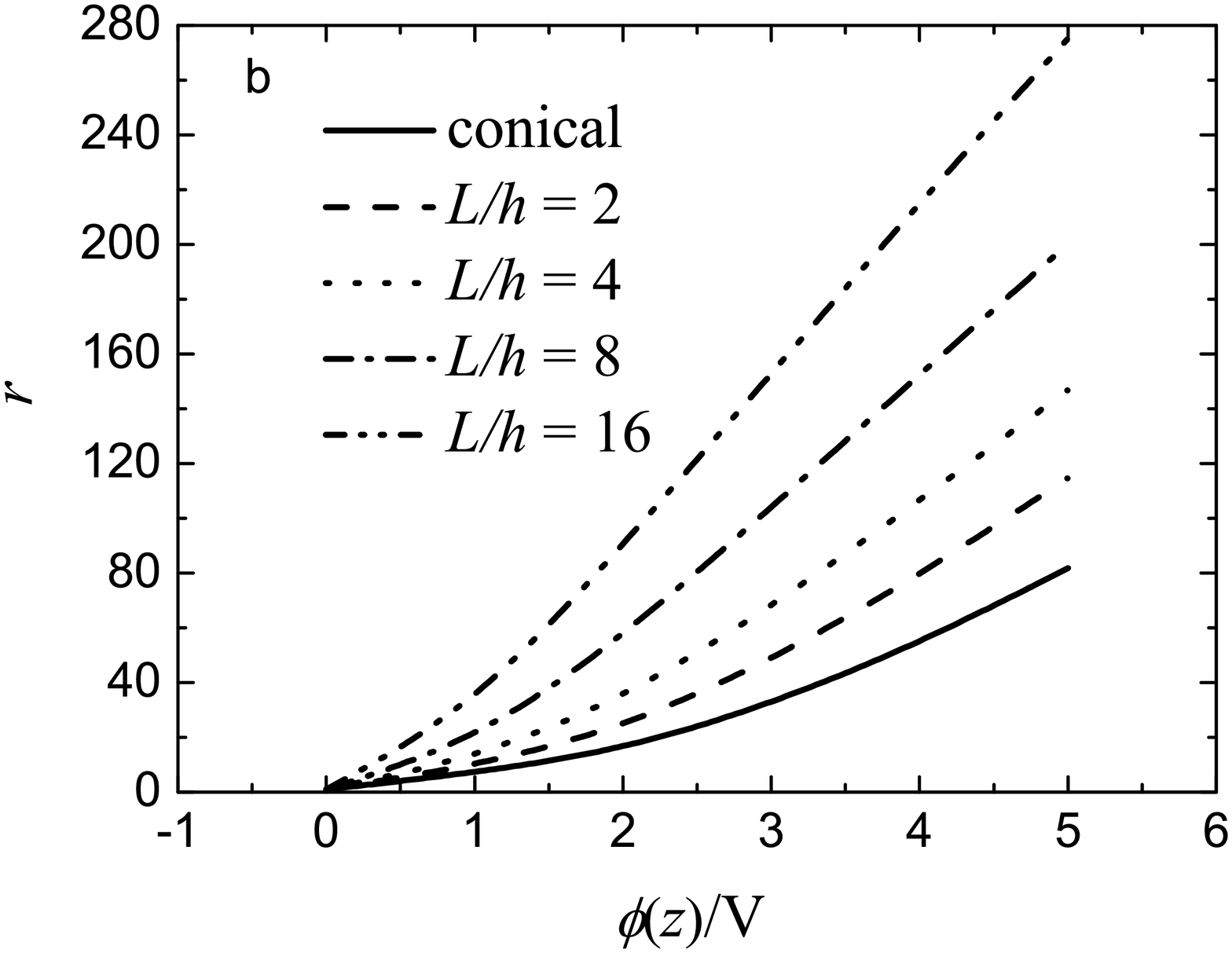}
\end{minipage}
\caption{(a) Ion current as a function of the applied potential and (b) corresponding ion current rectification as a function of absolute value of the applied potential.}
\label{IV}
\end{figure}

Figure 2a shows the current-voltage curves of the bipolar nanofluidic diode. The rectification behavior is observed in all cases. The currents increase with the increase of the applied potential due to the increase of electric field. The values of the reverse current increase significantly slower than that of the forward current, so the reverse currents seems nearly constant. What is also observed is that the forward current is higher at higher $d/h$, which means more tapered, at a given voltage. The phenomenon is more signification at higher voltage. The rectification degree is shown in Figure 2b. The rectification degree is defined as the ratio of the forward and reverse current values at a given voltage. In consistence with the current-voltage curves, the rectification degree increases with tapering. The rectification degree at voltage of $5V$ of the bullet-like nanopore of $d/h=16$ is about 3.5 times that of the rectification degree of conical nanopore. The current rectification degree of the dipolar nanofluidic diode based on different bulltlike nanopores is higher than that of unipolar nanofluidic diode. The enhancement of the rectification degree and ion current at forward bias is mainly due to the increase of the pore lumen with tapering.

\begin{figure}
\centering
\begin{minipage}[t]{0.3\textwidth}
    \centering
    \includegraphics[scale=0.25]{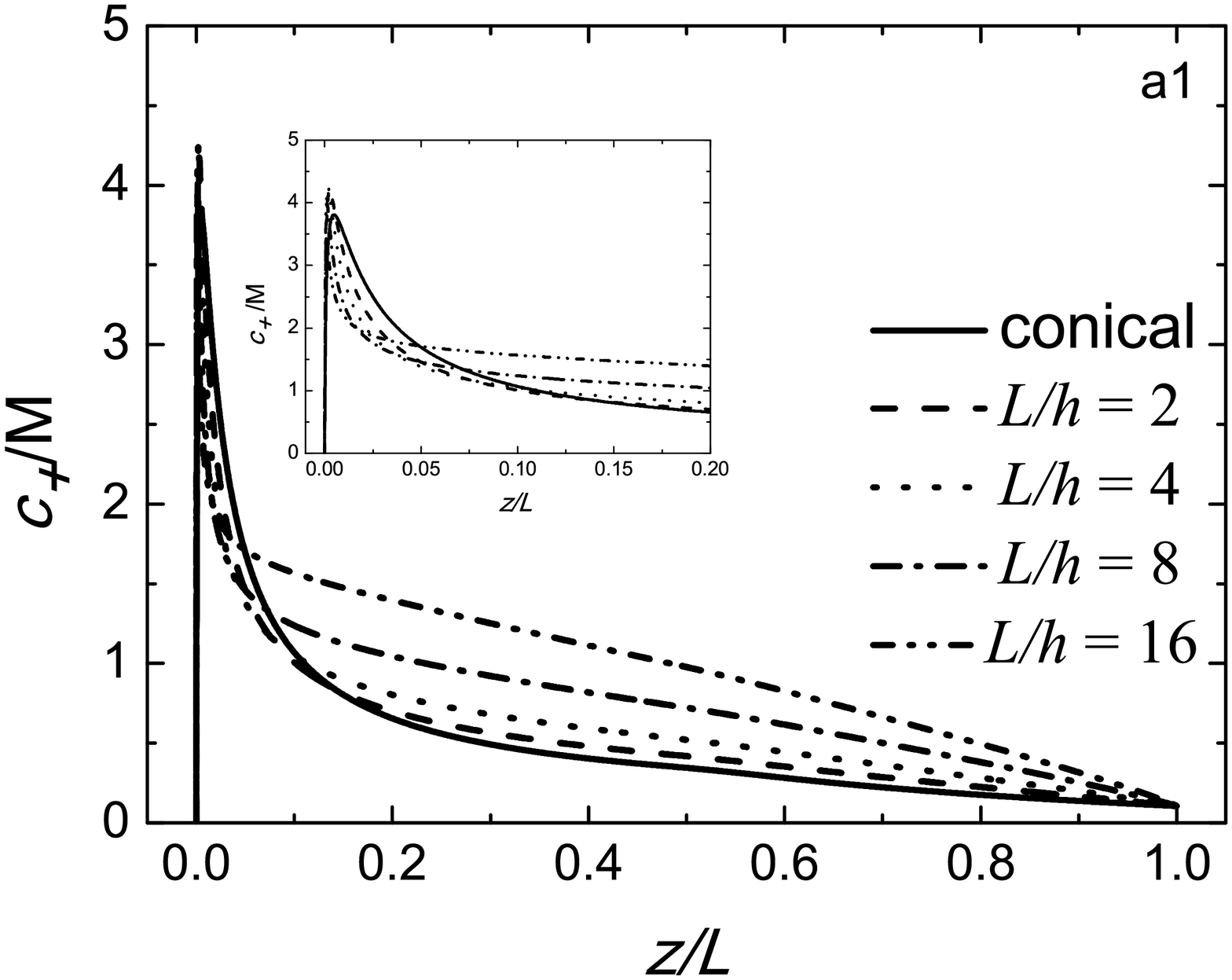}
\end{minipage}
\hspace{60pt}
\begin{minipage}[t]{0.3\textwidth}
    \centering
    \includegraphics[scale=0.25]{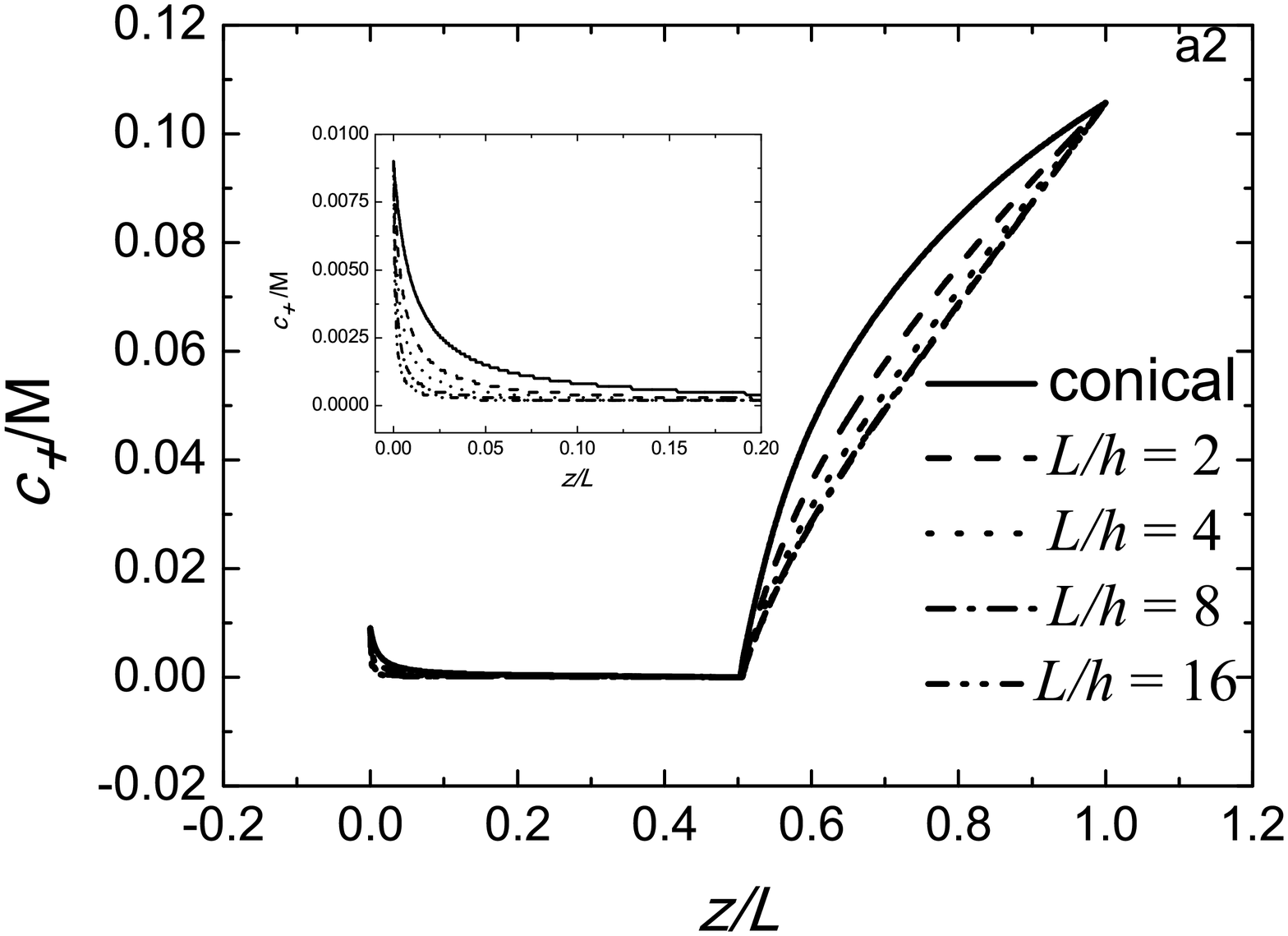}
\end{minipage}
\begin{minipage}[t]{0.3\textwidth}
    \centering
    \includegraphics[scale=0.25]{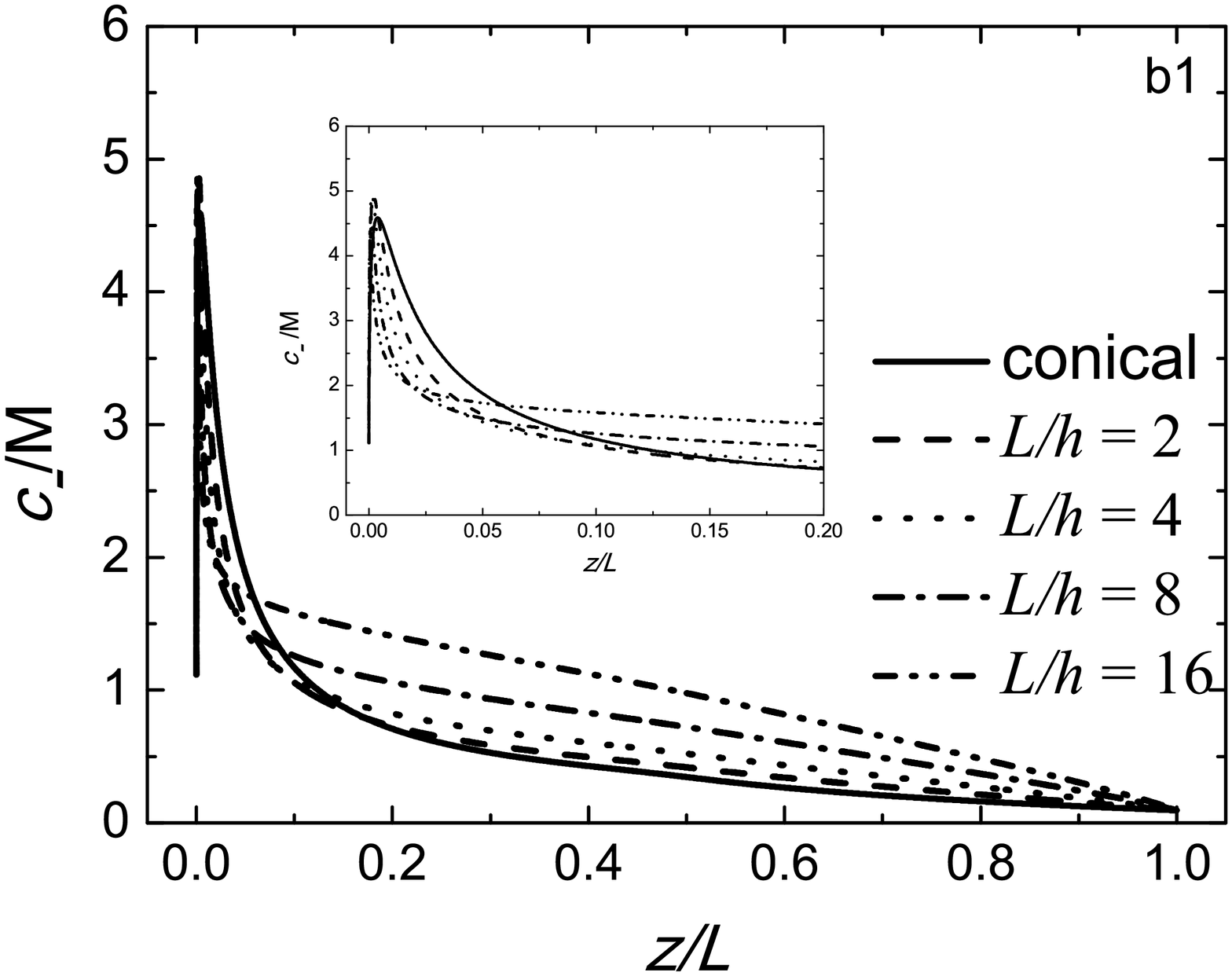}
\end{minipage}
\hspace{60pt}
\begin{minipage}[t]{0.3\textwidth}
    \centering
    \includegraphics[scale=0.25]{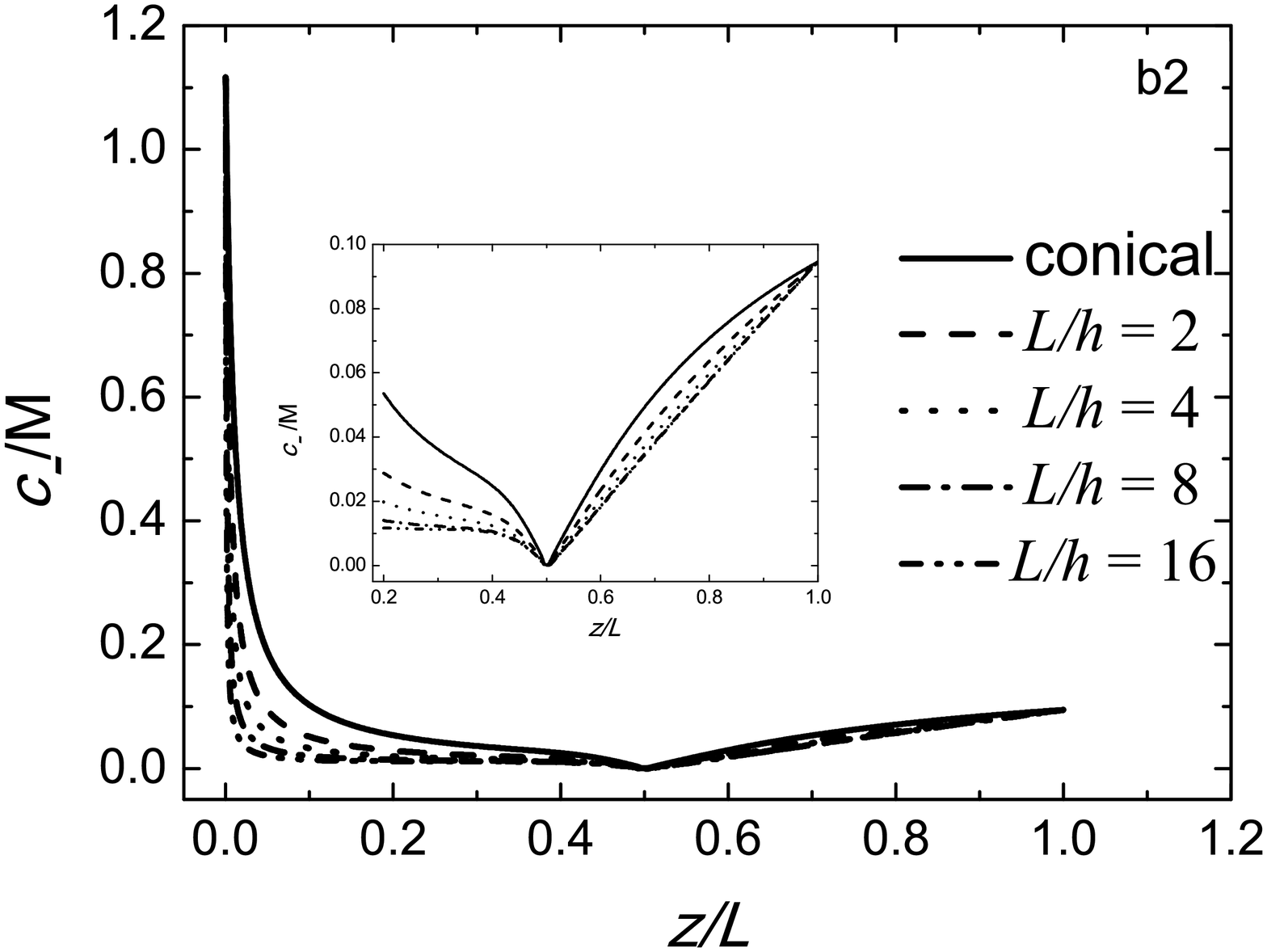}
\end{minipage}
\begin{minipage}[t]{0.3\textwidth}
    \centering
    \includegraphics[scale=0.25]{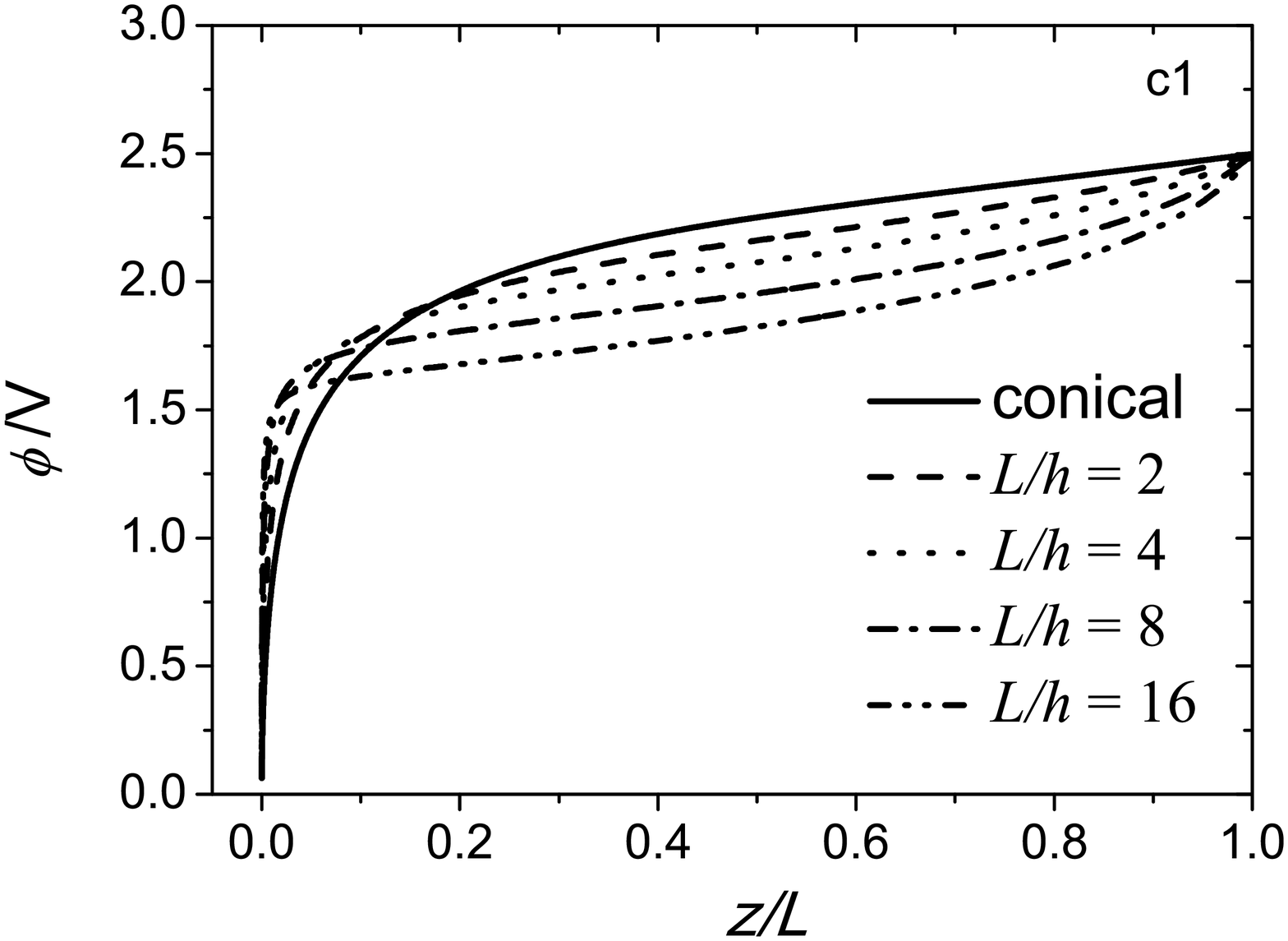}
\end{minipage}
\hspace{60pt}
\begin{minipage}[t]{0.3\textwidth}
    \centering
    \includegraphics[scale=0.25]{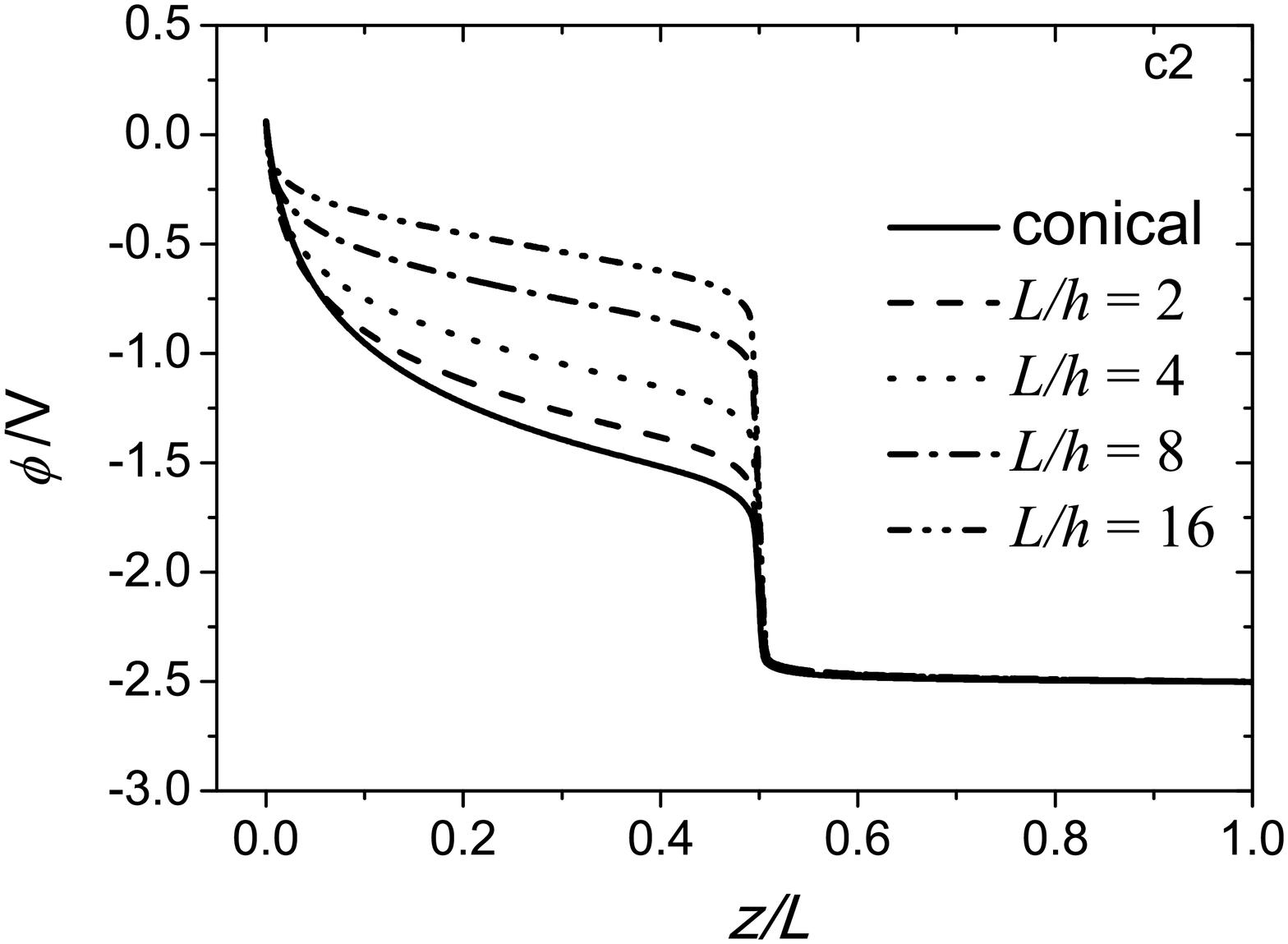}
\end{minipage}
\caption{Profiles of ion concentration and electric potential for nanopores of different shape. The figures with prefixes (a), (b) and (c) are for profiles of cation concentration, anion concentration and electric potential, respectively. The figures with suffixes (1) and (2) are for forward bias ($\phi_R=2.5V$) and reverse bias ($\phi_R=-2.5V$), respectively.}
\label{profile}
\end{figure}

Figure 3 presents the concentration profiles of the positive and negative ions and the electric potential at $2.5V$ and $-2.5V$ of various nanopores as in Figure 2. Figures 3(a1) and 3(b1) show the ion concentration profiles in forward bias at $\phi_R=2.5V$. Concentrations of both positive and negative ions show similar behaviors. The concentration profiles decrease along the nanopore from the left to the right. The ions accumulate significantly at the left openings with concentrations higher than the bulk concentration by an order. Quantitatively, at the left zone of the nanopore, the concentration of the negative ion is a little higher than that of the positive ion. The concentrations of both positive and negative ions in the pore near the tip decrease with tapering. This is because the double electrical layer interaction, which has significant effect on the ion concentrations near the tip of the pore, weakens with the tapering. The more tapered the pore tip is, the narrower the peak of the ion density profiles are. Far away from the tip along the pore, the concentrations increase with tapering because the lumen is wider for more tapered pores.

Ion concentration profiles in reverse bias at $\phi_R=-2.5V$ are shown in Figures 3(a2) and 3(b2), which are completely different from that in forward bias. There are almost no ion distributing in the middle region of the nanopore. This phenomenon is called ion depletion in semiconductor science. Ion depletion occurs because the oppositely charged ions move away from the junction of the surface coating charge for reverse bias. From the figures, we can see that the pore shape has no influence on the size of the depletion zone.

Figures 3(c1) and 3(c2) show the electric potential profiles along the nanopores of different shapes at forward and reverse bias respectively. In forward bias, the electric potential increases along the nanopore from the cathode to the anode. The electric field drives the positive ions to the pore tip and these ions accumulate there. The negative ions also tend to accumulate in the tip region to preserve the electroneutrality. So the density profiles of the positive and negative ions are similar. For the conical pore, the electric potential evolves smoothly along the pore. For the electric potential in the bulletlike pore, there is a jumpwise behavior near the tip, which explains the narrower peaks of the ion density profiles near the tip in the bulletlike pore. In reverse bias, as shown in Figure 3(c2), an abrupt drop occurs for the electric potential at the depletion zone. The electric field drags the oppositely charged ions away from the junction of the fixed charge of opposite signs on the inner surface of the pore. As a result, an ion depletion zone forms at the junction and a sharp variation in electric potential takes place. The abruptness increases with the tapering of the nanopore tip, which means higher electric field in the depletion zone and more effectively blocking the ionic current.

\section{Conclusion}
The Poisson-Nernst-Planck (PNP) equations were employed to investigate the effect of pore shape on ion current rectification in a bipolar nanofluidic diode. We have shown that the shape of the pore tip has great influence on the ion current rectification behavior. The bipolar nanofluidic diode with more tapered tip has higher ion current rectification degree and forward current because of effectively longer lumen. The results obtained here indicate that there may be optimal pore structure for the performance of the bipolar nanofluidic diode. By varying the nanopore radius equation we may get the structure-function relation and obtain the optimal structure.

\medskip
\noindent
\textbf{Acknowledgements}

The authors acknowledge the financial support of National Natural Science Foundation of China (NSFC) 20990234, 20874111, 20973176, 508211062, 973 Program of the Ministry of Science and Technology (MOST) 2011CB808502.




\medskip
\noindent
\textbf{References}
\bibliographystyle{model1-num-names}
\bibliography{bipolardiode}







\end{document}